**Dr. Mansur BEŞTAŞ**
https://orcid.org/0000-0002-8192-2044
Siirt / TÜRKİYE

# MERKEZİYETSİZ FİNANS (DeFi)

# DECENTRALIZED FINANCE (DeFi)

## ÖZET

Blok zinciri teknolojisinden gücünü alan merkezi olmayan finans alanı her geçen gün büyümektedir. Son birkaç yıl önce ortaya çıkmış olan bu alan, bugün 70 milyar dolarlık varlık yönetmektedir. Bu çalışmada, merkezi olmayan finans kavramı ele alınmış, geleneksel finanstan farkları açıklamıştır. Ardından yasal düzenlemeler ile uyumuna ve uyum sağlamak için gereksinimlere değinilmiştir. Merkezi olmayan finans alanının sunduğu finansal hizmetler ve bu hizmetleri sağlarken araç olarak kullandığı borsa ve stablecoinler hakkında değerlendirmede bulunulmuştur. Ekonomik etkileri, güvenlik ve gizlilik boyutları incelenmiştir. Çalışmada genel olarak yasal, ekonomik, güvenlik, gizlilik ve piyasa manipülasyonunu kapsayan merkezi ve merkezi olmayan finans arasındaki farkları sistematik olarak analiz edilmiştir. Merkezi ve merkezi olmayan finans hizmetleri arasında ayrım yapmak için yapılandırılmış bir metodoloji sunulmuştur.

## ABSTRACT

Decentralized finance, powered by blockchain technology, is growing day by day. This field, which emerged a few years ago, today manages $70 billion in assets. In this study, the concept of decentralized finance is discussed and explained the differences from traditional finance. Then, compliance with the legal regulations and the requirements to ensure compliance are mentioned. An evaluation has been made about the financial services offered by the decentralized finance field and the stock market and stablecoins that it uses as a tool while providing these services. Its economic effects, security and, privacy dimensions are examined. In the study, the differences between centralized and decentralized finance, which generally covers legal, economic, security, privacy, and market manipulation, are systematically analyzed. A structured methodology is presented to distinguish between centralized and decentralized financial services.
**Keywords:** decentralized finance, FinTech, financial regulation, blockchain, distributed ledger technology.

## 1. GİRİŞ

Finansın insanlık tarihi kadar eskiye gitmesiyle beraber bugün için hayatın her alanındadır. Merkezi finans tarihi Mezopotamya’ya kadar geri götürülebilmektedir. İnsanların günlük hayatta kullandıkları ürün ve malzemenin artması ile geniş çapta takas ihtiyacı ortaya çıkmıştır. Takas yöntemleri zaman içerisinde yetersiz kalmış ve değerinin herkes için aynı olduğu yeni meta kavramlar ortaya çıkmıştır. Bugün bu meta kavramları para olarak adlandırmaktayız. Para kavramı uzun bir süre fiziksel niteliği ile var olmuştur ancak teknolojinin gelişmesi ile dijital ortamda var olan genellikle kredi kartları ile günlük hayat ile etkileşime geçer hale gelmiştir. Sanal ortamda var olan para, merkezi kurumlar olan bankalar ve benzeri işletmeler aracılığı ile kayıt altına alınmış ve sahipleri için güvenliği sağlanmıştır.

Blok zincir teknolojisi ile paranın sahiplik ve güvenlik konuları yeniden yorumlanmıştır. Blok zincir teknolojisinin geliştirilmesi ile ortaya çıkan merkeziyetsizlik, dijital varlıkların kimseye güven ihtiyacı duymadan ve onayına ihtiyaç olmadan A noktasından B noktasına gidebilmesini sağlamıştır. Finansal teknolojiler ve piyasalar blok zincir teknolojisinin sağladığı hareket kabiliyetinden etkilenmiştir. Günümüzde, finans kavramı merkezi finans ve merkeziyetsiz (merkezi olmayan) finans olmak üzere ikiye ayrılmış durumdadır. Merkeziyetsiz finans, blok zincir teknolojisinin finans konusunda özelleşmiş yeni alt alanıdır. Merkezi finansın temel hizmetleri olan borç alma ve borç verme süreçleri blok zincir teknolojisine taşınmıştır.

Bu çalışmada, merkezi finans ve merkezi olmayan finans (Decentralized finance - DeFi) alanının sistematik olarak karşılaştırması yapılmıştır. Her iki alanın teknolojik olarak farklılıklarına değinilmiş, varlık yönetimine bakış açısı irdelenmiş ve kullanıcı açısından varlık yönetiminde avantaj ve dezavantajlar değerlendirilmiştir. DeFi alanının yasal düzenlemeler açısından güncel durumu hakkında bilgi verilmiş olup sahip olduğu fırsatlar açısından değerlendirmesi yapılmıştır.

## 2. FİNANSIN KAVRAMSAL ÇERÇEVESİ ve GELİŞİMİ

Finans, birey, kurum ve toplumların kendilerine ait olan mali kaynakları tahsis etme ve yönetme yöntemlerini inceleyen bilim dalı olmakla beraber bankacılık, gayrimenkul ve yatırım yönetimi gibi alanları da kapsamaktadır. Finansın temel araştırma konularından biri, finansal kararların kaynakların tahsisini etkileme biçimi ve yatırımların değerlemesi ve riskini ne şekilde etkilediğini bulmaktır. Finans alanında bu temel araştırma konusuna cevap bulabilmek amacıyla finans uzmanları, finansal tablolar, bilançolar, gelir tabloları ve nakit akış tablolarının da dâhil olduğu finansal verileri analiz etmek ve anlamak amacıyla çeşitli araç ve teknikler kullanmaktadır (Rostow, 1974).

Finansın temel ilkelerinden biri, paranın değerinin enflasyon ve diğer faktörler nedeniyle zaman içinde değiştiğini belirten paranın zaman değeridir. Paranın zaman değeri önemlidir çünkü finansal kaynakların gelecekte belirli bir finansal hedefe ulaşmak için belirli bir zaman diliminde ne kadar para ayırmaları gerektiğinin anlaşılmasına yardımcı olur (Mokyr, 1992).

Finansın diğer bir temel ilkesi risk ve getiridir. Tüm yatırımlar doğaları gereği risk taşımaktadır ve bir yatırımın potansiyel getirisi genellikle barındırdığı risk seviyesi ile ilişkilidir. Finans biliminde, yatırımların risk ve getiri seviyelerini belirlemek amacıyla finansal modelleme, tarihsel verilerin analizi ve olasılık teorisi dâhil olmak üzere çeşitli yöntemler kullanılmaktadır (Acemoğlu, Johnson ve Robinson, 2005:402-405).

Finans; para ve varlıkların yönetimi ilgilenen bir alan olmasından dolayı insanlık tarihinde çok eski zamanlardan beridir var olmaktadır. İnsanlık tarihinde borç alma, borç verme, yatırım yapma, bütçeleme ve tahmin yapma gibi birçok alanda kendini göstermiştir. İnsanların her daim kaynaklarını yönetme ihtiyacı olduğundan dolayı finans tarihi ve insan toplumunun gelişimi iç içe geçmiş bir şekilde ilerlemiştir.

Bilinen en eski finans sistemleri, Mezopotamya ve Mısır'da ortaya çıkmış en eski uygarlıklara kadar izlenebilir. Bu toplumlar, bilinen para birimlerinden farklı olarak mal ve hizmet takasına dayalı "meta parası" olarak bilinen para birimi kullanmıştır. Antik Yunan ve Roma'da, madene dayalı paranın yaygınlaşması ve basımının gelişmesi ile birlikte faizle borç verme kavramı, finans biliminin daha da ilerlemesinin sağlamıştır. Katolik kilisesi ortaçağ süresince, dini tarikatlar ve manastırlar aracılığı ile büyük arazilerin ve finansal kaynakların yönetilmesinde önemli rol oynamıştır. Rönesans döneminde ticari bankacılığın gelişimi muhasebe uygulamalarının gelişime neden olmuştur. Modern muhasebenin temelleri modern finansın temellerinin atılmasına yardımcı olmuştur (Miller, 1999).

Modern çağda, finans alanı, teknolojinin gelişim ve ekonomik faaliyetlerin çeşitlenmesi ve küreselleşmesi ile gelişimini devam ettirmiştir. Günümüzde finans, global işletmelerin, hükümetlerin ve bireylerin işleyişinin merkezinde bulunmaktadır (Das, 2006).

Merkezi finans (Centralized finance – CeFi), bir organizasyon içinde bulunan finansal kaynakların tek merkezden yönetim ve kontrol biçimini ifade eder. CeFi, finansal planlama ve karar verme süreçlerinin koordinasyonunun ve finansal politika ve prosedürlerinin ortaya çıkarılmasının ve uygulamaya konulmasını da içermektedir.

CeFi, bir kuruluşun finansal varlıkları ve operasyonel faaliyetleri açısından istikrarını ve sürdürülebilirliğini sağlamada anahtar rol oynamaktadır. Kuruluşlar, finansal süreçlerini merkezileştirmek suretiyle kaynaklarını daha etkin bir şekilde kullanabilmek ve kötü yönetimden kaynaklı oluşan riskleri azaltabilmektedir. CeFi'nin bir diğer faydası kuruluşların daha yüksek verimlilik ve maliyet etkinliği elde etmesidir. Kuruluşlar, finansal süreçleri ve operasyonel faaliyetlerinin standartlaştırarak manuel müdahale ihtiyacının azaltabilmektedir. Bu durum, finansal performansın daha etkin izlenebilmesi ve kuruluş genelindeki yönelimlerin belirlenmesi, kuruluşların finansal problemleri daha hızlı belirlemesine ve değerlendirmeye almasına yardımcı olmaktadır. CeFi, finansal verilerin ve süreçlerin değerlendirilmesi ve değerlendirmenin desteklenmesi amacıyla güncel teknolojiden faydalanarak teknolojik ve altyapısal yatırımlara ihtiyaç duymaktadır (Purdy, Langemeier ve Featherstone, 1997).

Genel olarak değerlendirildiğinde, CeFi planlama, kontrol, görünürlüğü geliştirmek isteyen kuruluşlar için değerli bir yaklaşım sunmaktadır. Ancak CeFi yaklaşımının faydaları uygulamaya alınmadan önce potansiyel faydaları ve zorlukları detaylı olarak değerlendirmeye tabi tutulmalıdır.

### 2.1. Finansal Teknolojiler

Finans alanının teknolojik gelişmelerden etkilenmesi kaçınılmaz bir durumdur. Finans alanı teknolojik gelişmeleri kullanırken teknolojide kendine has bir alan oluşturmuştur. Bilişim teknolojilerinin finansal alanda kullanılması Fintech (financial technologies) olarak anılmaktadır. Fintech genel olarak finansal teknoloji, finansal hizmetlerin geliştirilmesi için kullanılan bir terimdir. Online bankacılık, mobil ödeme sistemleri, yatırım yönetimi ve kripto para birimlerinin kapsayan tüm uygulama ve teknolojileri tanımlamaktadır (Mention, 2019).

Fintech, finans alanında yenilikçi bir değişim oluşturmasının temel yollarından biri, tüketiciler için erişilebilirliği ve rahatlığı arttırmasıdır. Örnek olarak, çevrimiçi bankacılık ve mobil ödeme sistemleri, bireylerin finansal işlemlerinin mekândan ve zamandan bağımsız olarak sadece internet erişimi ile herhangi bir yerden yönetmelerini sağlar. Bu durum, özellikle finansal hizmetlerin yetersiz olarak iletilebildiği topluluklarda, geleneksel finansal kuruluşlara erişimi kısıtlı olan veyahut bankaları fiziksel olarak ziyaret etme imkânının zor olan bireylere kayda değer fayda sağlamıştır (Bollaert , Lopez-de-Silanes ve Schwienbacher, 2021; Safiullah ve Paramati, 2022).

Fintech, finansal kuruluşlara sağladığı faydaların yanında ilgili kuruluşların çalışma şeklini de değiştirmektedir. Birçok banka ve finansal kuruluşlar, süreçlerinin ve operasyonlarını kolaylaştırmak, maliyetleri azaltmak ve müşteri hizmetlerini daha verimli hale getirmek amacıyla finansal teknoloji çözümlerini kullanmaktadır. Fintech kullanımı finansal verileri analiz etmek ve en az hata ile karar vermek için yapay zekâ ve makine öğrenmesi algoritmalarının kullanımını kapsamaktadır. Genel olarak değerlendirildiğinde, Fintech, finans alanının ve ilgili sektörün gelişimi ve değişiminde önemli rol oynamaktadır ve bu durumun giderek artan bir etki ile devam etmesi muhtemel olarak görülmektedir (Rabbani, Khan ve Thalassinos, 2020; Vučinić, 2020).

### 2.2. Blok Zincir Teknolojisi

Blok zinciri teknolojisi, işlemleri kaydetmek ve gerçekliğini doğrulamak için kriptolojik tekniklerden ve eşler arası iletişimden faydalanan merkezi olmayan ve dağıtılmış dijital bir defterdir. İlk olarak 2008 yılında teorik olarak bahsedilen bu teknoloji daha sonra Bitcoin para biriminin hayata geçirilmesi ile uygulamalı olarak ortaya konulmuştur (Nakamoto, 2008). Blok zinciri teknolojisi, verileri tarihsel bir zincir yapısında birbirine bağlı veri bloklarında depolayarak çalışmaktadır. Verilerin bloklarda saklanmasından dolayı "blok zinciri" adı verilmektedir. Her veri bloğu, ağda gerçekleşen işlemlerin güvenli ve değiştirilemez bir kaydının bir zaman damgası ile saklanması ve önceki bloğa ait bir bağlantı içerir. Blok zincir teknolojisi doğasında merkezi olmama durumu bulunmaktadır. Blok zincir teknolojisinin merkezi olmaması, merkezi bir otoriteden bağımsız, kurcalamaya ve sansüre karşı dirençli hale getirir. Bu özellikler finans, tedarik zinciri ve oylama sistemleri gibi çeşitli alanlar için dikkat çekici hale getirir. Blok zincir teknolojisinin ortaya koyduğu en önemli faydaların başında güvenli ve şeffaf işlem süreçlerinin kolaylaştırmasıdır. Blok zincir teknolojisi merkezi olmaması ve işlemleri şeffaf bir biçimde ele almasından dolayı aracılara ihtiyaç duymadan süreçlerin doğrulanmasına olanak tanımaktadır, dolandırıcılık ve istismar riskini azaltır ve verimliliği arttırmaktadır. Genel olarak değerlendirildiğinde, blok zincir teknolojisi veri, işlemlerin kaydedilmesi için güvenli ve şeffaf bir çözüm sunmaktadır. Blok zincir teknolojisinin sağladığı çözüm aracı ihtiyacına sahip birçok endüstride ciddi anlamda değişikliğe neden olabilecek potansiyel bir teknolojidir. Merkezi olmayan teknolojisi, kurcalamaya ve sansüre karşı dirençli oluşu çeşitli uygulamalarda kullanımı açısından çekici bir teknoloji haline getirmektedir (Nakamoto, 2008).

## 3. MERKEZİ OLMAYAN FİNANS

Blok zincir teknolojisinin birçok sektörde kullanımı artmıştır. Özellikle blok zincir teknolojisinin akıllı sözleşmeler ile süreçleri programlanabilir ve otonom hale getirmesi ile birçok sektörde uygulamalar ortaya çıkmaya başlamıştır. Blok zincir teknolojisinin kullanımının ortaya çıktığı sektörlerden biri de finans alanıdır. Blok zincir teknolojisinin finans alanında kullanılmaya başlanması ile finans sektörü yeniden yorumlanmaya ve adlandırılmaya başlanmıştır. Geleneksel finans ve kuruluşları CeFi olarak adlandırılmaya başlanmıştır. Blok zincir teknolojisi kullanılan finans uygulamaları ise merkezi olmayan finans olarak adlandırılmaya başlanmıştır.

DeFi blockchain teknolojisi üzerine kurulu ve bankalar veya finansal kurumlar gibi merkezi aracılara ihtiyaç duymadan çalışan finansal sistemleri ifade eder. Bu merkezi olmayan yaklaşım, geleneksel finansal sistemlere kıyasla daha fazla şeffaflık, erişilebilirlik ve güvenlik sağlar (Buterin, 2014).

DeFi sistemleri, taraflar arasındaki anlaşmanın süreçlerinin ve şartlarının akıllı sözleşme üzerine yazıldığı, kendi kendini yürüten akıllı sözleşmeleri kullanmaktadır. Finansal süreçler, akıllı sözleşmeler ile otomatik yürütülmekte ve aracıların süreci kolaylaştırması ihtiyacını ortadan kaldırmaktadır (Ethereum White Paper, 2014).

DeFi, bankalar veya finansal kurumlar gibi aracılara ihtiyaç duymadan borç verme, borç alma, ticaret ve ödemeler gibi finansal işlemleri ve hizmetleri kolaylaştırmak için blockchain teknolojisini kullanmaktadır (Liu, 2020). DeFi, bireylerin ve kuruluşların, açık kaynaklı yazılıma dayalı ve merkezi olmayan otonom kuruluşlar (DAO'lar) tarafından yönetilen merkezi olmayan platformlar ve protokoller aracılığıyla finansal hizmetlere doğrudan erişmesine olanak tanımaktadır (Park, Kim ve Lee, 2020). En bilinen DeFi platformları Ethereum, Solano, Cordano, Poligon, Avalanche, Llama, SushiSwap, Aave, Cake, Polkadot, Eos, Flow, ChainLink, Tron Defi'dir.

DeFi alanı gittikçe gelişen bir alandır. Bu alanda yeni projeler ortaya çıkmakta ve projelerin barındırdığı para miktarı gittikçe artmaktadır. Defillama sitesi üzerinde istatistikleri verilmiş olan 1445 adet DeFi projesinde kilitli varlık miktarı (Total Value Locked-TVL) toplamda 04/01/2023 tarihi 39.52 milyar dolara ulaşmıştır. Şekil 1'de ilk 10, sonraki 100 ve geri kalan projelerin toplam TVL değerleri verilmiştir.

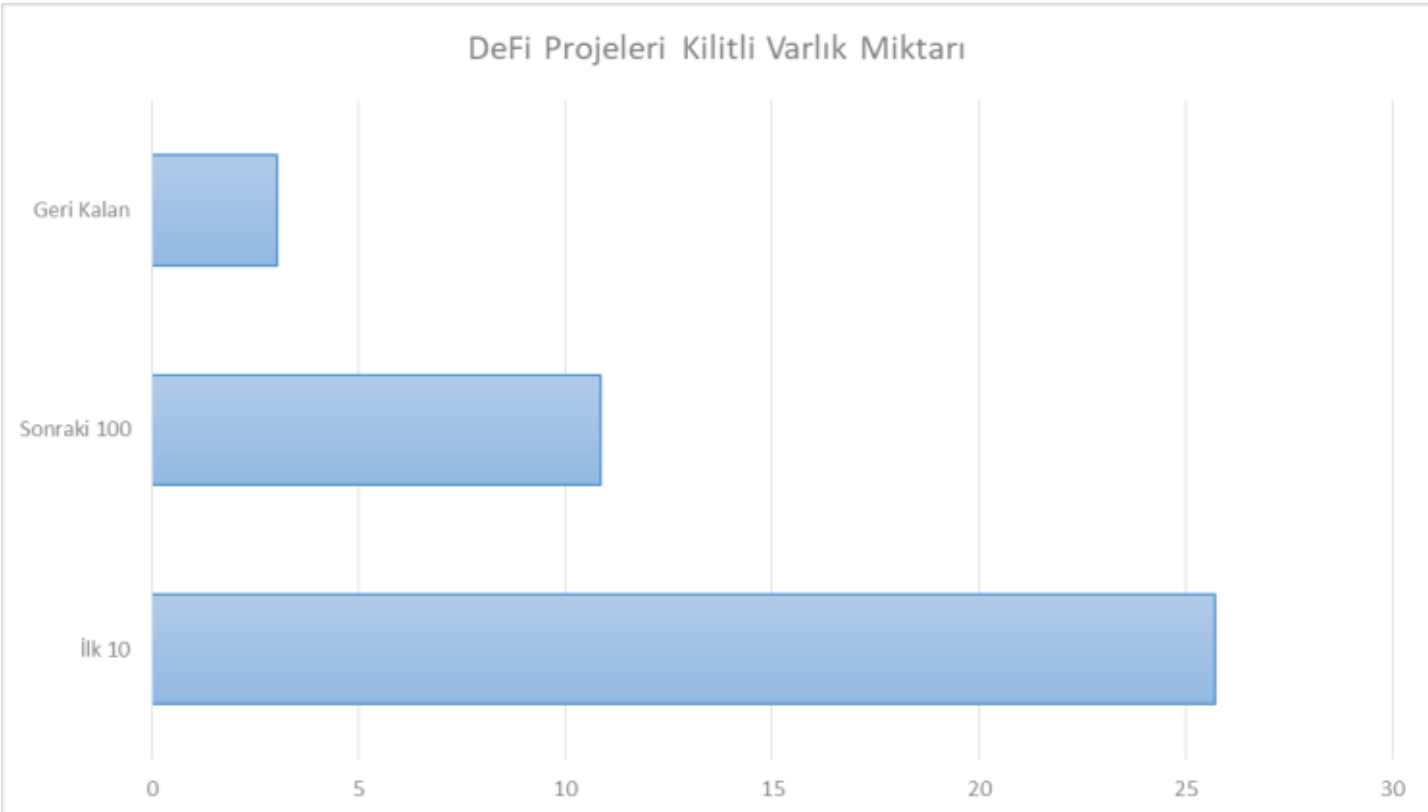

**Şekil 1. DeFi Projeleri Kilitli Varlık Miktarı, Kaynak**: URL1

Blok zincir teknolojinin sağladığı imkanlar ile bir çok girişimci kendi blok zincir ağını oluşturmuştur. Blok zincirler arasında temel olarak mutabakat protokolü, blok boyutu, Turing makinası teknolojisi gibi farklılıklar olsa da temelde blok zincir üzerinde güvenlikten ödün vermeden ölçeklenebilirlik problemlerine çözüm üretmektir. Blok zincirler bu nedenle çeşitli olmaktadır. Blok zincir teknolojisine dayalı oluşturulmuş zincirler kendi üzerinde faaliyet göstermek için akıllı sözleşme imkanı sunmaktadır. Bu nedenle bir blok zincir üzerinde birden fazla proje bulunabilmektedir. Şekil 2'de blok zincir üzerinde faaliyet gösteren farklı DeFi projelerinin toplam TVL değerinin oransal olarak ne kadarını barındırdığını göstermektedir.

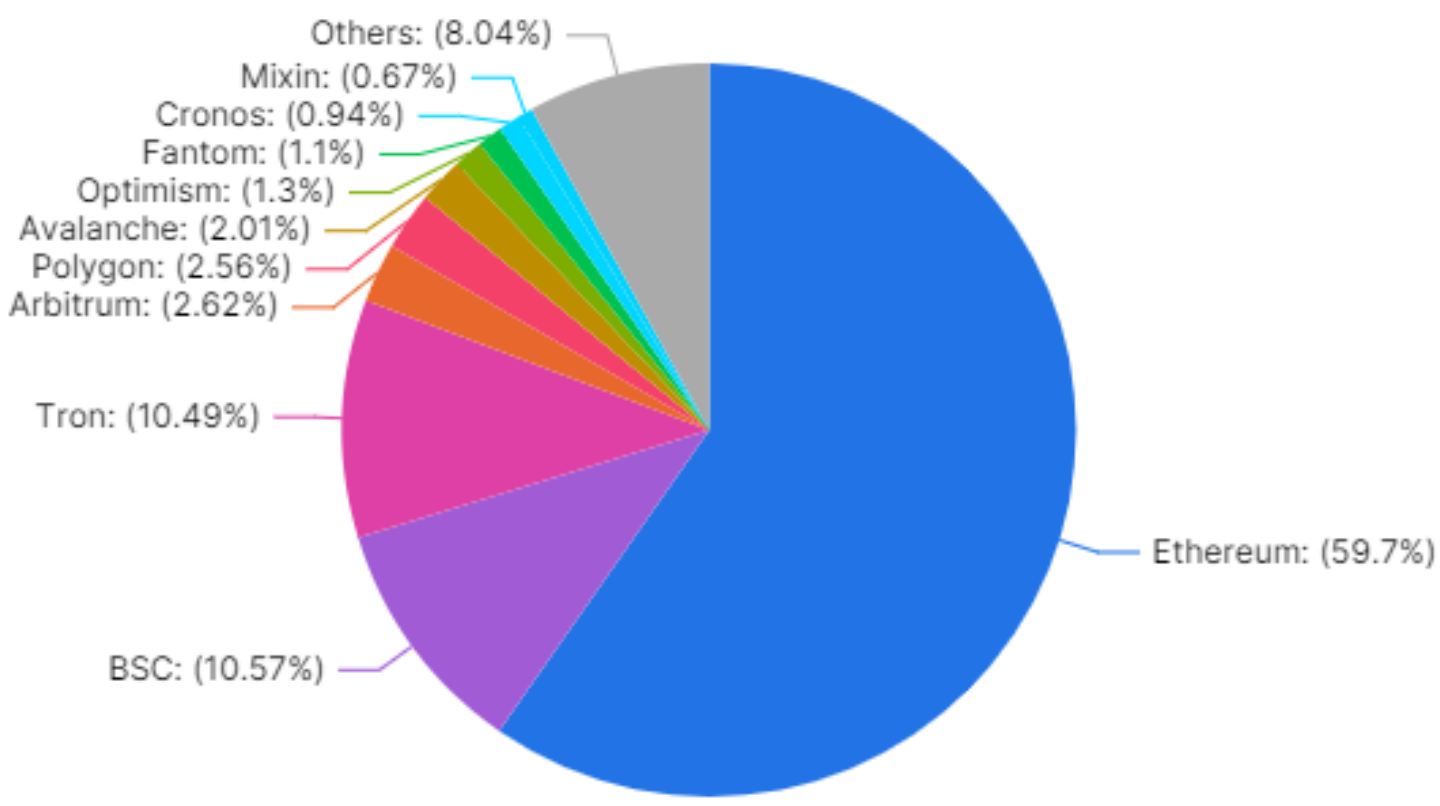

**Şekil 2. Blok Zincir Bazında Kilitli Varlık Oranları, Kaynak**: URL1

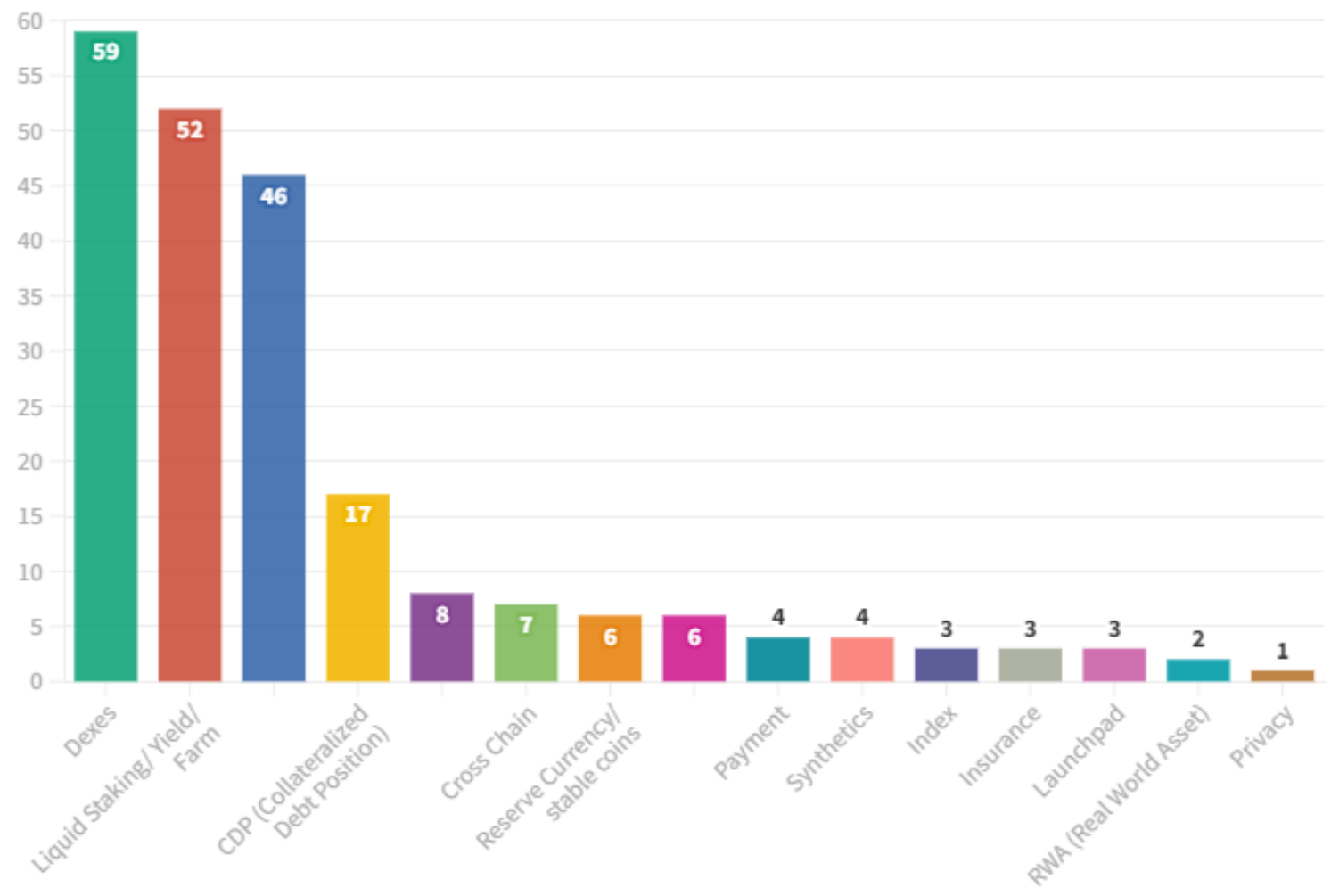

**Şekil 3. En Büyük 200 Defi Projesinin Kullanım Amacı, Kaynak:** URL2

DeFi projeleri borç verme ve borç alma gibi finansal ihtiyaçlar için çözümler üretmesinin yanında blok zincir teknolojisinin kendisi ile birlikte getirdiği ödeme, zincirler arası aktarım için köprüler, likidite sağlama gibi alanlar içinde kullanılmaktadır. Şekil 3'te başlıca sektörlere göre TVL değerleri görülmektedir.

DeFi'nin en belirgin faydalarından biri, bireylerin ve işletmelerin fiziksel bir banka hesabına veya finansal bir geçmişe ihtiyaç duymadan finansal hizmetlere erişiminin sağlanmasıdır. Geleneksel finansal süreçlerden bağımsız olarak ilerlemesinden dolayı daha geniş finansal kapsayıcılığa izin vermektedir (Buterin, 2014). DeFi, özellikle geleneksel finansal sistemlere sınırlı erişimi olan veya finansal geçmişleri nedeniyle finansal hizmetlerden mahrum kalmış olan ülkelerdeki bireyler için önemlidir. DeFi sistemleri genellikle, geleneksel finansal sistemlere kıyasla daha fazla şeffaflık ve daha düşük ücretler sağlayan eşler arası (P2P) temelinde çalışmaktadır (Buterin, 2014). Bunun nedeni, P2P işlemlerinin finansal işlemlerle ilişkili maliyetleri önemli ölçüde arttırabilecek aracılar gerektirmemesidir.

DeFi birçok fayda sağlamasına rağmen, halen değerlendirilmesi gereken bazı zorluklar bulunmaktadır. DeFi sistemleri için en temel zorluklardan biri kullanıcılar ve yatırımcılar için genel geçer bir finansal çerçevenin oluşturulmamış olmasıdır (Buterin, 2014). DeFi'nin likidite sorunlarının yanında düzenleyici belirsizliği problemleri bulunmaktadır (Liu, 2020).

DeFi kavramı yeni olduğu kadar ortaya çıkan sistemler de yenidir ve yeni olmasından dolayı sistemlerin kendi içinde barındırdığı siber güvenlik problemlerine karşı savunmasız olabilmektedir (Buterin, 2014). DeFi geliştiricilerinin ve kullanıcılarının, DeFi'nin uzun vadeli sürdürülebilirliğini ve benimsenmesini sağlamak için bu zorlukları dikkate alması ve bunları ele alacak önlemleri uygulaması önemlidir (Park vd., 2020).

DeFi, finansal erişebilirliği arttırmasının yanında bireyler ve işletmeler için maliyetleri azaltmak için önemli bir potansiyel sunmaktadır.

### 3.1. DeFi Özellikleri

DeFi, blok zincir gibi merkezi olmayan bir ağ üzerinde çalışan ve bankalar veya finansal kurumlar gibi aracılara ihtiyaç duymadan işlemleri ve finansal hizmetleri kolaylaştırmak için akıllı sözleşmelerden yararlanan bir finansal sistemdir. DeFi'yi benzersiz kılan ve geleneksel finanstan ayıran birkaç temel özelliği vardır.

**Ademi merkeziyetçilik**: DeFi, merkezi bir otorite tarafından kontrol edilmek yerine merkezi olmayan bir ağ üzerinde çalışır. İşlemleri doğrulayan ve yürüten düğümler olarak bilinen merkezi olmayan bir bilgisayar ağı tarafından yönetilir. Bu, tek bir arıza noktası olmadığı ve kullanıcıların sistemle aracı olmadan doğrudan etkileşime girebileceği anlamına gelir. Bu merkezi olmayan kontrol, DeFi'nin sansüre dayanıklı olmasını ve geleneksel finansal kurumlardan bağımsız olarak çalışabilmesini sağlar (Hertig, 2022).

**Şeffaflık**: DeFi işlemleri şeffaf ve değişmez bir deftere kaydedilerek kullanıcıların bir finansal varlığın veya sözleşmenin geçmişini görmesini kolaylaştırır, bu da tüm işlemlerin herkes tarafından görülebileceği anlamına gelir. Bu şeffaflık, kullanıcıların işlemlerin gerçekliğini doğrulamasına olanak tanır ve DeFi'nin şeffaf ve hesap verebilir olmasını sağlar (Royal, 2022).

**Erişilebilirlik**: DeFi, konumu veya mali durumu ne olursa olsun internet bağlantısı olan herkes tarafından erişilebilir. Bu, geleneksel finansal hizmetlere sınırlı erişimi olan ülkelerdeki insanların küresel ekonomiye katılmasını mümkün kılar ve DeFi'yi geleneksel finansal hizmetlere erişimi olmayan bireyler ve topluluklar için daha erişilebilir hale getirir (Hertig, 2022).

**Gözaltı dışı**: DeFi, işlemleri kolaylaştırmak için akıllı sözleşmeler kullanır; bu, kullanıcıların varlıklarını elde tutmak için aracılara güvenmeleri gerekmediği anlamına gelir. Bunun yerine, kullanıcılar dijital cüzdanları kullanarak kendi varlıklarını tutabilir ve kontrol edebilir. Bu gözetim dışı model, güvenliği artırır ve dolandırıcılık veya hırsızlık riskini azaltır (Royal, 2022).

**Güvenlik**: DeFi sistemleri, kriptografik protokollerle korunur ve bu da onları dolandırıcılığa ve bilgisayar korsanlığına karşı dirençli hale getirir (Schär, 2021).

**Programlanabilirlik**: DeFi sistemleri, alıcı ve satıcı arasındaki anlaşmanın şartlarının doğrudan kod satırlarına yazıldığı kendi kendini yürüten sözleşmeler olan akıllı sözleşmeler üzerine kuruludur. Bu, finansal süreçlerin otomasyonunu sağlayarak aracılara olan ihtiyacı azaltır ve verimliliği artırır (Schär, 2021).

**Birlikte çalışabilirlik**: DeFi sistemleri, zincirler arası finansal uygulamaların oluşturulmasına ve varlıkların likiditesinin artırılmasına izin vererek diğer blok zincir ağlarıyla etkileşime girebilir (Caldarelli ve Ellul, 2021).

Sonuç olarak DeFi, onu benzersiz ve yenilikçi bir finansal sistem yapan birkaç temel özelliğe sahiptir. Merkezi olmayan kontrolü, gözetimsiz modeli, şeffaflığı ve erişilebilirliği, onu geleneksel finansa uygun bir alternatif haline getirmekte ve finansal hizmetlerin sunulma ve erişilme biçiminde devrim yaratma potansiyeline sahiptir.

### 3.2. DeFi Yasal Uyumluluk Problemleri ve Çözümleri

DeFi, her ne kadar blok zincir teknolojisi sayesinde daha verimli ve şeffaf bir finansal hizmetleri sunmaya başlamış olsa dahi uzun vadede genele yayılan uygulanabilir çözümleri sunmak için bir dizi yasal uygunluk sorunu ile karşı karşıyadır.

DeFi, karşı karşıya kaldığı en önemli yasal problemlerden biri genel bir düzenleme çerçevesinin olmamasıdır. Düzenleme çerçevesinin olmaması potansiyel kullanıcıların hizmetleri kullanmasında belirsizlik yaratabilmekte ve ana akım olarak benimsenmesini engellemektedir. DeFi paydaşları, hükümetlerden ve sektörün düzenleyici kurumlarından düzenlemeye yönelik netlik ve rehberlik çağrısında bulunmuştur (Schueffel, 2021).

DeFi için bir diğer uyumluluk problemi kara para aklama ve terörün finansmanı ilgili risklerdir. DeFi'nin merkezi olmayan mimarisi bireylerin yasal olmayan faaliyetlerde bulunmasını kolaylaştırmaktadır. DeFi, platformlarında, dolandırıcılık ve piyasa manipülasyonu potansiyeli bulunmaktadır. DeFi platformları piyasa manipülasyonu tespit etmek ve önlemek için piyasa gözetimi ve ticaret raporlaması gibi önlemler uygulaması gerekmektedir.

DeFi'de bulunan bu problemler, müşterinizi tanıyın (Know your customer – KYC) ve şüpheli faaliyet raporlaması (Suspicious activity reports – SAR) gibi etkili kara para aklamayla mücadele (Anti-money laundering – AML) ve terörün finansmanıyla mücadele (Counter-terrorism financing – CTF) önlemlerinin uygulanması ile aşılabilecektir. Bir diğer uyumluluk problemi, DeFi'de tüketici haklarının ihlal riskidir. DeFi yeni gelişen bir alan olmasından dolayı ve geleneksel finans süreçlerinden farklı bir yol izlemesinden dolayı geleneksel tüketici koruma düzenlemelerine tabi değildir, bu durum tüketicilerin CeFi alanında sahip olduğu korumaya sahip olmamasına neden olmaktadır (OECD, 2022).

Genel olarak değerlendirildiğinde, DeFi'nin uzun vadede kullanılmasının ve genel olarak kabulünün sağlanması için yasal gereklilikler olan KYC, AML ve CTF önlemlerinin alınması gerekmektedir. Geliştiriciler ve kullanıcıların da dahil olduğu DeFi paydaşlarının birlikte çalışması ve yasalara mümkün olduğunca uyumlu bir ekosistemi oluşturması gerekmektedir.

### 3.3. DeFi ve CeFi Karşılaştırması

DeFi ile CeFi'nin karşılaştırılması, bu alanda yatırım yapacak bireyler ve kurumlar için önemli bir konudur. Bu çalışmada, bu iki finansal sistemi birbirinden ayırt etmek için kullanılabilecek temel faktörler ana hatlarıyla açıklanmıştır.

DeFi ve CeFi arasındaki temel farklardan biri, ademi merkeziyet seviyeleridir. DeFi sistemleri merkezi değildir, yani tek bir varlık veya varlık grubu tarafından kontrol edilmezler. Bunun yerine, sisteme katılan ve sistemin yönetimine katkıda bulunan bir kullanıcı ağına güvenirler. Öte yandan CeFi sistemleri merkezileştirilmiştir, yani bir banka veya finans kurumu gibi tek bir kuruluş veya kuruluşlar grubu tarafından kontrol edilmektedir (Zohar ve Schinasi, 2019).

DeFi ve CeFi arasındaki bir diğer önemli fark, akıllı sözleşmeleri kullanmamalarıdır. DeFi sistemleri, alıcı ve satıcı arasındaki anlaşmanın şartlarının doğrudan kod satırlarına yazıldığı kendi kendini yürüten sözleşmeler olan akıllı sözleşmelere dayanır. Akıllı sözleşmeler, işlemlerin yürütülmesi gibi süreçlerin otomasyonuna izin verir ve sistem içindeki tüm faaliyetlerin şeffaf ve değişmez bir kaydını sağlar. CeFi sistemleri tipik olarak akıllı sözleşmeler kullanmaz (Dos Santos vd., 2022).

DeFi ve CeFi arasında ayrım yapmak için kullanılabilecek üçüncü bir faktör, operasyonlarındaki şeffaflık düzeyidir. DeFi sistemleri, merkezi olmayan ağlara ve şeffaf akıllı sözleşmelere dayandıkları için genellikle CeFi sistemlerinden daha şeffaftır. CeFi sistemleri ise genellikle operasyonlarında şeffaflıktan yoksundur (Katona, 2021).

Dikkate alınması gereken dördüncü bir değişken, güvenlik ve risk düzeyidir. DeFi platformları, operasyonlarını denetleyecek merkezi bir otorite olmadığı için merkezi olmayan yapıları nedeniyle güvenlik risklerine karşı daha savunmasız olabilir. CeFi platformları ise merkezi yapıları ve düzenleyici gözetime tabi olmaları nedeniyle daha güvenli olabilir.

**Tablo 1. Defi ve CeFi Karşılaştırması**

| Özellik | CeFi | DeFi |
|---|---|---|
| Hizmetler | Alım Satım, Borç verme, itibari paradan kripto paraya döndürme, ödeme, borç verme | Alım Satım, Borç verme, ödeme, borç verme |
| Merkezilik | Kripto ticareti merkezidir (Binance, Kraken, Coinbase). | Kripto para ticareti merkezi değildir (UniSwap, PancakeSwap). |
| Kullanıcı Yetkisi | Kullanıcılar KYC yaptırmak zorundadır ve kara para aklama düzenlemelerinden etkilenir. | Kullanıcı kaydı yaptırmak zorunda değildir. Yasal düzenlemelerden etkilenmezler. |
| Güven | Kripto ve para yönetimi ve korunması platform yetkisindedir | Cüzdan kontrolü bireydedir. Süreçler akıllı kontratların belirlediği standartlarda gerçekleşir. |
| Müşteri destek | Platformlar belli bir düzeyde müşteri hizmeti sağlar. | Müşteri desteği yoktur. |
| Şeffaflık | Kaynakları ve cüzdanlar arası aktarım şeffaf değildir. | Açık kaynak kodludur. Cüzdanlar arası aktarım açıktır. |
| Esneklik | Oldukça esnektir. İtibari paralar ile takas gerçekleştirilebilir | Esnek değildir. İtibari para desteği yoktur. |
| Zincirler arası aktarım | İtibari para ve kripto varlıklar zincirler arası aktarımı rahatlıkla yapılır. | Zincirler arası aktarım zincirlerin desteklediği köprülerin yetenekleri ile sınırlıdır. Zincirler arası aktarım teknik bilgi gerektirir. |

**Kaynak**: URL3

Sonuç olarak, DeFi ve CeFi sistemleri, ademi merkeziyet seviyelerine, akıllı sözleşmelerin kullanımına ve şeffaflık seviyelerine göre farklılaştırılabilir. DeFi sistemleri daha fazla merkezi olmayan ve şeffaflık sunarken, CeFi sistemleri daha merkezidir ve operasyonlarında genellikle şeffaflıktan yoksundur.

### 3.4. DeFi'nin En Yaygın Özellikleri

DeFi uygulamalarının geleneksel finansal kurumlara veya aracılara dayanmadığı, bunun yerine blockchain teknolojisi gibi merkezi olmayan ağlarda çalışmaktadır.

En yaygın DeFi özelliklerinden biri, akıllı sözleşmelerin kullanılmasıdır. Akıllı sözleşmeler, alıcı ve satıcı arasındaki anlaşmanın şartlarının doğrudan kod satırlarına yazıldığı kendi kendini yürüten sözleşmelerdir. Bu sözleşmeler, şeffaf ve güvenli finansal işlemlere izin veren blockchain teknolojisi kullanılarak kolayca uygulanabilir ve doğrulanabilir (Nakamoto, 2008).

Bir başka yaygın DeFi özelliği, merkezi olmayan borsaların (DEX) kullanılmasıdır. DEX'ler, merkezi bir otoriteye ihtiyaç duymadan kripto para varlıklarının eşler arası ticaretine izin vererek, geleneksel merkezi borsalara kıyasla daha fazla güvenlik ve daha düşük ücretler sağlar (Buterin, 2014).

DeFi uygulamaları ayrıca, kullanıcıların aracı olarak geleneksel finansal kurumlara ihtiyaç duymadan varlıkları ödünç vermesine ve ödünç almasına olanak tanıyan merkezi olmayan borç verme ve ödünç alma platformlarını da sıklıkla kullanır. Bu platformlar genellikle, borçluların bir krediyi güvence altına almak için varlıkları teminat olarak göstermesi gereken teminatlı krediler kullanmaktadır (Buterin, 2014).

Diğer bir yaygın DeFi özelliği, ABD doları gibi gerçek dünyadaki bir varlığın değerine sabitlenmiş dijital varlıklar olan stablecoin'lerin kullanılmasıdır. Bu stablecoin'ler, kripto para piyasasında daha fazla istikrar ve öngörülebilirlik sağlayarak onları DeFi uygulamaları için çekici bir seçenek haline getirmektedir (Buterin, 2014).

Genel olarak, akıllı sözleşmeler, DEX'ler, merkezi olmayan borç verme ve ödünç alma platformları ve stablecoin'ler gibi DeFi mülkleri, finans sektörünü büyük ölçüde etkiledi ve geleneksel finans kurumlarını bozma potansiyeline sahiptir.

### 3.5. DeFi Hizmetleri

DeFi, merkezi olmayan platformlarda çalışan çok çeşitli finansal hizmetler sunan, kripto para birimi ve blok zincir endüstrisinde hızla büyüyen bir sektördür. En yaygın DeFi hizmetleri arasında merkezi olmayan borsalar (DEX), istikrarlı paralar, borç verme ve ödünç alma platformları ve merkezi olmayan özerk kuruluşlar (DAO) bulunur.

Merkezi olmayan borsalar (DEX), kullanıcıların merkezi bir otoriteye veya aracılara ihtiyaç duymadan kripto para birimleri ve diğer dijital varlıkları alıp satmalarına olanak tanır. DEX'ler, blockchain ağları üzerinde çalışarak kullanıcıların kendi varlıkları üzerinde kontrol sağlarken birbirleriyle doğrudan ticaret yapmalarını da sağlar. DEX'ler, hızlı ve güvenli ticaret sağlama yeteneklerinin yanı sıra merkezi borsalara (CEX'ler) kıyasla daha fazla likidite ve daha düşük ücretler sunma potansiyelleri nedeniyle giderek daha popüler hale gelmiştir. Dune Analytics tarafından hazırlanan bir rapora göre DEX'ler, 2020'nin 2. çeyreğinde toplam DeFi işlem hacminin %60'ından fazlasını oluşturmaktadır (Consensys Analytics, 2020).

Stablecoin'ler, ABD doları gibi gerçek dünyadaki bir varlığın değerine sabitlenmiş dijital varlıklardır. Stablecoin'ler, değişken kripto para birimi piyasasında istikrarlı ve güvenli bir değer deposu sağlamak üzere tasarlanmıştır ve bu da onları özellikle DeFi uygulamaları için kullanışlı kılar. En popüler stablecoin'ler arasında Tether (USDT), DAI ve USDC bulunur. CoinMarketCap tarafından hazırlanan bir rapora göre, stablecoin'lerin toplam piyasa değeri 2020'nin 2. çeyreğinde 10 milyar doları aşmıştır (CoinMarketCap, 2020).

Ödünç verme ve ödünç alma platformları, kullanıcıların eşler arası işlemler yoluyla kripto para birimini ve diğer dijital varlıkları ödünç vermesine veya ödünç almasına olanak tanır. Bu platformlar, çeşitli kredi koşulları ve faiz oranları sunarak kullanıcıların borçlanma veya borç verme stratejilerini özelleştirmelerine olanak tanır. Borç verme ve borç alma platformları, DeFi ekosisteminde giderek daha popüler hale geldi ve DeFi borç verme platformlarında kilitli toplam değer (TVL) 2020'nin 2. çeyreğinde 5 milyar doları aşmıştır (DeFi Pulse, 2020).

Merkezi olmayan özerk kuruluşlar (DAO'lar), kullanıcıların akıllı sözleşmeler aracılığıyla toplu olarak kararlar almasına ve eylemler gerçekleştirmesine olanak tanıyan merkezi olmayan yönetişim sistemleridir. DAO'lar genellikle açık kaynaklı yazılım geliştirme veya hayır kurumları gibi merkezi olmayan projeleri yönetmek ve finanse etmek için kullanılır. DAO'lar, DeFi DAO'lardaki TVL'nin 2020'nin 2. çeyreğinde 1 milyar doları aşmasıyla DeFi projelerini yönetmenin ve finanse etmenin popüler bir yolu haline gelmiştir (DeFi Pulse, 2020).

Sonuç olarak, merkezi olmayan platformlarda çalışan çok çeşitli finansal hizmetler sunan DEX'ler, stablecoin'ler, borç verme ve ödünç alma platformları ve DAO'lar en yaygın DeFi hizmetleridir. Bu hizmetler, DeFi ekosisteminde önemli ölçüde benimsendi ve bu hizmetlerde kilitlenen toplam değer milyarlarca dolardır.

### 3.6. CeFi ve DeFi Borsaları

CeFi ve DeFi borsaları, farklı prensipler üzerinde çalışan ve farklı amaçlara hizmet eden iki farklı kripto para borsasıdır. CeFi, bankalar ve borsalar gibi merkezi bir otorite tarafından düzenlenen geleneksel finansal kurumları ifade eder. Bu kurumlar, katı düzenleyici yönergeleri izledikleri ve müşteri fonlarını korumak için ileri teknoloji kullandıkları için güvenlikleri ve güvenilirlikleri ile tanınırlar. Bununla birlikte, merkezi olmayan muadillerine göre daha yavaş ve daha pahalı olma eğilimindedirler. CeFi borsaları Merkezi Borsa (CEX) olarak adlandırılmaktadır (Aspris ,Foley, Svec ve Wang, 2021).

DeFi, blockchain teknolojisi ve akıllı sözleşmeler kullanarak merkezi olmayan bir platformda çalışan finansal sistem ve hizmetleri ifade eder. DeFi borsaları merkezi değildir, yani merkezi bir otorite tarafından kontrol edilmezler ve işlemleri kolaylaştırmak için bir kullanıcı ağına güvenirler. Bu, daha hızlı, daha ucuz ve daha şeffaf finansal işlemlere izin verir, ancak sistemi düzenleyecek merkezi bir otorite olmadığı için daha yüksek riskleri de beraberinde getirir. DeFi borsaları Merkezi Olmayan Borsa (DEX) olarak adlandırılmaktadır (Heimbach, Wang ve Wattenhofer, 2021).

CeFi ve DeFi borsaları arasında güvenlik, şeffaflık, erişilebilirlik ve düzenleyici gözetim dâhil olmak üzere birkaç önemli fark vardır. CeFi borsaları, merkezi yapıları ve yasal uyumlulukları nedeniyle genellikle daha güvenli ve güvenilir olarak kabul edilirken, DeFi borsaları, merkezi olmayan yapıları nedeniyle daha fazla şeffaflık ve erişilebilirlik sunar. Bununla birlikte, kullanıcı fonlarını koruyacak merkezi bir otorite olmadığı için DeFi borsaları, bilgisayar korsanlığı ve diğer güvenlik risklerine karşı da daha savunmasızdır. En önemli CeFi borsaları; Binance, Kraken, Coinbase iken DeFi borsaları olarak UniSwap, 1inch ve PancakeSwap örnek verilebilir (Barbon ve Ranaldo, 2021).

### 3.7. DeFi ve CeFi Platformlarında Borç Verme ve Alma

CeFi ve DeFi, kripto para piyasasında borç verme ve borç alma konusunda iki farklı yaklaşım bulunmaktadır. CeFi, borç verme ve borç alma hizmetleri sunan geleneksel finans kurumlarını ifade ederken, DeFi, bu hizmetleri kolaylaştırmak için blockchain teknolojisini kullanan merkezi olmayan platformları ifade eder. CeFi üzerinde borç verme ve alma sürecinde finansal kurumlar borç alan ve veren arasında aracı görevi görmektedir. Borç almak isteyenler, kurumlardan kredi başvurusunda bulunmakta ve verilecek kredi, kredi talebinde bulunanların kredi itibarına ve krediyi geri ödeme kabiliyetine göre onaylanır veya reddedilir. Borç verenler paralarını CeFi kuruluşlarına yatırabilmekte ve faiz şeklinde getiri elde edebilmektedir (Rosic, 2020).

CeFi kuruluşları, devlet kurumları tarafından yasal düzenlemelere tabi olmasından dolayı risk az olmaktadır. Bu nedenle CeFi kuruluşlarının borç verme ve alma süreçleri daha güvenli ve istikrarlıdır. Ancak süreçlerin birden fazla onay işlemine tabi tutulması daha yüksek ücretler ve kredi onay sürecinin şeffaflığın olmaması gibi dezavantajları da beraberinde getirir (Heimbach vd., 2021).

DeFi kuruluşları, borç alma ve verme süreçlerini kolaylaştırmak için akıllı sözleşmeler kullanan merkezi olmayan platformlarda çalışmaktadır. Bu platformlar, yüksek düzeyde şeffaflık ve güvenlik sağlayan blockchain teknolojisi üzerine kuruludur.

DeFi borç verme ve borç almanın bir avantajı, marj ticareti ve eşler arası borç verme dahil olmak üzere daha geniş bir finansal hizmet yelpazesine erişebilme yeteneğidir. Ayrıca, kredi veya coğrafi kısıtlamalar nedeniyle geleneksel finansal kurumlara erişimi olmayan kişilere daha fazla erişilebilirlik sağlar. Bununla birlikte, düzenleme eksikliği ve bilgisayar korsanlığı veya dolandırıcılık potansiyeli nedeniyle DeFi borç verme ve ödünç almanın genellikle daha riskli olduğu kabul edilir (Barbon ve Ranaldo, 2021).

Sonuç olarak, CeFi ve DeFi borç verme ve borç alma, kripto para piyasasındaki finansal hizmetlere farklı yaklaşımlar sunar. CeFi daha fazla istikrar ve güvenlik sunar, ancak daha yüksek ücretler ve daha az şeffaflıkla gelirken, DeFi daha fazla erişilebilirlik ve daha geniş bir hizmet yelpazesi sunar, ancak genellikle daha riskli olarak kabul edilir.

### 3.8. CeFi ve DeFi Stablecoin'leri

Stablecoin'ler genellikle bir teminat mekanizması kullanarak belirli bir varlığa veya para birimine karşı istikrarlı bir değer sağlamayı amaçlayan bir tür kripto para birimidir. CeFi stablecoin'leri, bankalar veya diğer finansal kurumlar gibi merkezi kurumlar tarafından çıkarılır ve yönetilirken, DeFi stablecoin'leri merkezi olmayan ağlar ve protokoller tarafından çıkarılır ve yönetilir. CeFi stabilcoin örnekleri arasında Tether (USDT) ve USDC bulunur. DeFi stablecoin'lerine örnek olarak DAI ve sUSD verilebilir (Zapotochnyi, 2022).

Tether (USDT) gibi istikrarlı sabit para birimleri, merkezi bir otorite tarafından verilir ve desteklenir ve genellikle ABD doları gibi geleneksel itibari (Fiat) para birimlerinin değerine sabitlenir. Bu stablecoin'ler tipik olarak geleneksel finansal altyapı aracılığıyla çıkarılır ve yönetilir, bu da onları düzenleyici incelemeye ve potansiyel güven ve şeffaflık sorunlarına karşı savunmasız hale getirir (Wei, 2018).

Öte yandan, MakerDAO'nun Dai'si gibi istikrarlı para birimleri, merkezi olmayan protokoller ve akıllı sözleşmeler aracılığıyla blockchain ağında yayınlanır ve yönetilir. Bu stablecoin'ler tipik olarak kripto para birimleri ve emtialar gibi çeşitli varlıklar tarafından desteklenir ve merkezi bir otorite tarafından kontrol edilmez. Sonuç olarak, DeFi stablecoin'leri daha fazla şeffaflık ve merkezsizleştirme sunar, ancak merkezi olmayan yönetişime ve olası akıllı sözleşme güvenlik açıklarına dayanmaları nedeniyle ek riskler de oluşturabilir (Schirrmacher, Jensen ve Avital, 2021).

Benimseme ve kullanım söz konusu olduğunda, hem CeFi hem de DeFi stablecoin'leri kripto para piyasasında büyük ilgi görüyor. Bir CoinMarketCap raporuna göre, Tether şu anda piyasa değerine göre üçüncü en büyük kripto para birimi ve stablecoin olarak ilk sırada bulunmaktadır. Tether'i USD coin, Binance USD ve Dai takip etmektedir. Kullanıma sunulmuş olan toplam 134 adet stablecoin bulunmakta olup, Pazar büyüklüğü 138 milyar dolardır. Pazarın büyük bir kısmını Şekil 4'te görüldüğü üzere ilk dört coin oluşturmaktadır (CoinMarketCap, 2021).

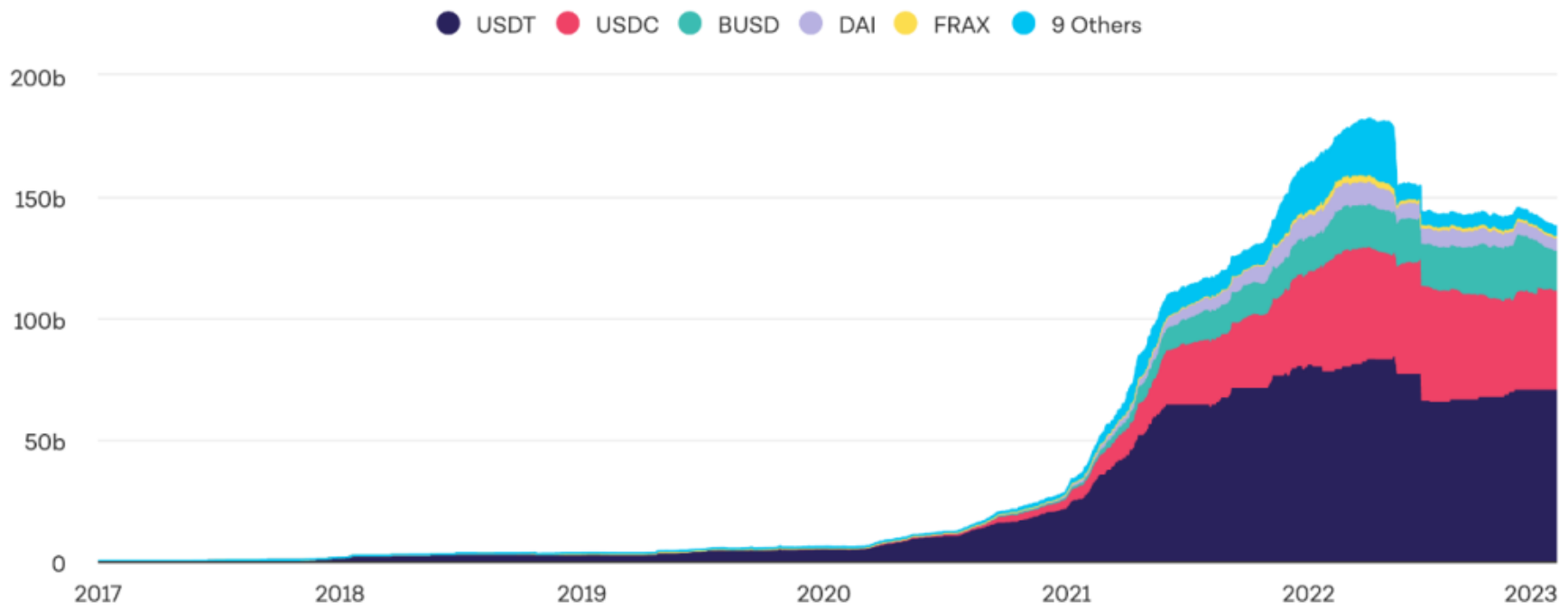

**Şekil 4. Stablecoin Pazar Büyüklüğü, Kaynak:** URL4

Hem CeFi hem de DeFi stablecoin'lerinin güçlü ve zayıf yönleri vardır. CeFi stablecoin'leri, köklü finansal kurumlar tarafından desteklendikleri için genellikle güvenilirdir ve geniş çapta kabul görmektedir. Ancak, aynı zamanda merkezileştirme risklerine ve potansiyel düzenleyici müdahale ve manipülasyona da tabidirler. Öte yandan, DeFi stablecoin'leri, blockchain teknolojisi üzerine inşa edildikleri ve akıllı sözleşmelerle yönetildikleri için daha fazla merkezi olmayan ve şeffaflık sunar. Düzenleyici gözetim eksikliği, daha az güvenilirlik ve kabul görmeye yol açabilmektedir.

Genel olarak, CeFi ve DeFi stablecoin'leri arasında seçim yapmak, nihayetinde bireysel risk toleransına ve merkezi veya merkezi olmayan yönetişim tercihine bağlıdır. Her iki stabilcoin türünün de kendi güçlü ve zayıf yönleri vardır ve bireylerin bir karar vermeden önce riskleri ve faydaları dikkatlice tartması önemlidir.

### 3.9. DeFi Sisteminin Ekonomik Etkileri

DeFi, merkezi bir aracıya ihtiyaç duymadan finansal hizmetler sağlamayı amaçlayan kripto para piyasasının hızla büyüyen bir sektörüdür. DeFi protokolleri ve platformları, borç verme, borç alma ve ticaret gibi çeşitli finansal hizmetler için merkezi olmayan çözümler sağlamak üzere blockchain teknolojisini kullanmaktadır.

DeFi'nin en önemli yönlerinden biri akıllı sözleşmelerin kullanılmasıdır. Akıllı sözleşme, alıcı ve satıcı arasındaki hüküm ve koşulların doğrudan kod satırlarına yazıldığı, kendi kendini yürüten bir sözleşmedir. Finansal işlemlerin otomatik ve şeffaf bir şekilde yürütülmesini ve karmaşık finansal süreçlerin otomasyonunu sağlar. Ancak DeFi'nin de kendi ekonomik dinamikleri ve potansiyel güvenlik açıkları vardır. Ana endişelerden biri, piyasa manipülasyonu potansiyelidir. Piyasa manipülasyonu, bir varlığın fiyatını yapay olarak etkilemek için çeşitli taktiklerin kullanılması anlamına gelir, örneğin: sahte haber veya ticari aklama kullanımı.

DeFi, akıllı sözleşmeler ve diğer merkezi olmayan uygulamalar (dApp'ler) tarafından piyasa manipülasyonu yaşayabilir. Örneğin, kötü niyetli bir aktör, belirli bir varlığın fiyatı belirli bir seviyeye ulaştığında çok sayıda satış emrini otomatik olarak gerçekleştiren ve ani bir fiyat düşüşüne neden olan bir akıllı sözleşme oluşturabilir.

DeFi'nin bir başka potansiyel güvenlik açığı, işlemlerin verimli bir şekilde yürütülmesini sağlayan kullanıcı tarafından sağlanan sermaye havuzları olan likidite havuzlarına dayanmasıdır. Likidite havuzlarının tükenmesi, likidite krizlerine ve piyasa istikrarsızlığına yol açabilir.

DeFi'de, bilgisayar korsanlarının Harvest Finance'in likidite havuzundan 20 milyon dolar çekmek için akıllı sözleşmelerdeki güvenlik açıklarından yararlandığı 2020 olayı gibi likidite havuzu manipülasyonu vakaları olmuştur. Bu sorunları ele almak için DeFi protokolleri ve platformları, aşağıdakiler de dahil olmak üzere çeşitli karşı önlemler uygulamıştır. Genel olarak, DeFi ekonomisi ve operasyonları, endüstri istikrarını ve güvenliğini sağlamak için daha fazla araştırma ve geliştirme gerektiren karmaşık konulardır (Gogo, 2020).

### 3.10. CeFi ve DeFi Enflasyonu

CeFi, bankalar ve geleneksel borsalar gibi merkezi otoriteler tarafından kontrol edilen geleneksel finansal sistemleri ifade eder. DeFi, merkezi olmayan ve merkezi kurumlar yerine bilgisayar ağları tarafından kontrol edilen finansal sistemleri ifade eder. Hem CeFi hem de DeFi enflasyon yaşayabilir, ancak enflasyonun arkasındaki mekanizmalar ve faktörler her sistem için farklıdır.

CeFi sistemindeki enflasyon genellikle para arzındaki artıştan veya paranın değerindeki düşüşten kaynaklanır. Merkez bankası gibi merkezi bir otorite, daha fazla para basarak veya yeni yaratılan parayla varlık satın alarak para arzını artırabilir. Bu para arzı artışı, mal ve hizmetlere olan talep para arzı ile aynı oranda artmazsa enflasyona neden olabilir. Enflasyon, ekonomik istikrarsızlık veya para birimine olan güven kaybı gibi faktörler nedeniyle bir para birimi değer kaybettiğinde de ortaya çıkabilir (Bank of England, 2021; Cao, 2015).

DeFi sistem enflasyonu genellikle yeni belirteçlerin çıkarılmasından veya mevcut belirteçlere yönelik artan talepten kaynaklanır. DeFi sistemlerinde, jetonlar (token) genellikle ağa katılım için bir ödül olarak veya sistemin gelişimini finanse etmenin bir yolu olarak verilir. Dolaşımdaki madeni para sayısı, bu madeni paralara olan talepten daha hızlı arttığında enflasyon meydana gelebilir. DeFi sistemlerinde, token talebi arzdan daha hızlı büyüdüğünde de enflasyon meydana gelebilir, bu da token fiyatını artırabilir (Qin, Zhou, Afonin, Lazzaretti ve Gervais, 2021).

Özetle, CeFi sistemleri para arzındaki artış veya para birimlerinin değer kaybetmesi nedeniyle enflasyon yaşayabilir ve DeFi sistemleri yeni jetonların çıkarılması veya mevcut jetonlara olan talebin artması nedeniyle enflasyon yaşayabilir.

CeFi, merkezi bir platformda çalışan geleneksel bir finansal sistemi ifade eder. CeFi sistemlerinde finansal işlemler, banka veya finans kurumu gibi merkezi bir otorite tarafından işlenir ve doğrulanır. CeFi sistemleri yüksek oranda düzenlenir ve kısmi rezerv bankacılığı gibi geleneksel finansal ilkeleri takip eder.

CeFi planının ana özelliklerinden biri, borç verme yoluyla yeni para yaratma yeteneğidir. Bir banka bir bireye veya işletmeye borç para verdiğinde, yeni para yaratır ve bunu borçlunun hesabına ekler. Bu süreç "kısmi rezerv bankacılığı" olarak bilinir ve bankaların sıfırdan yeni para yaratmasına olanak tanır (Cao, 2015).

Yeni para birimlerinin oluşturulması CeFi sisteminde enflasyona yol açabilir. Enflasyon, para arzının mal ve hizmet talebinden daha hızlı artmasıyla oluşur. Sonuç olarak, paranın değeri düşer ve mal ve hizmetlerin fiyatı yükselir. Bu, bireylerin ve işletmelerin satın alma gücünü azaltabilir ve yaşam standartlarını düşürebilir.

DeFi ise blockchain platformunda çalışan merkezi olmayan bir finansal sistemi ifade eder. DeFi sistemleri merkezi bir otorite tarafından kontrol edilmez, bunun yerine işlemleri doğrulamak ve işlemek için merkezi olmayan düğümlerden oluşan bir ağa dayanır. DeFi sistemleri, internet bağlantısı olan herkesin kullanımına açık olduğu için şeffaflığı, güvenliği ve erişilebilirliği ile bilinir.

DeFi sisteminin ana özelliklerinden biri, ABD doları gibi belirli bir varlığa sabitlenmiş dijital varlıklar olan stablecoin'ler oluşturma yeteneğidir. Stablecoin'ler, diğer dijital varlıklarla aynı dalgalanmalara tabi olmadıkları için DeFi sistemlerindeki oynaklığı azaltmak için kullanılabilir.

Stablecoin'in oluşturulması, DeFi sistemlerinde deflasyona yol açabilir. Deflasyon, para arzı mal ve hizmet talebinden daha hızlı düştüğünde meydana gelir. Sonuç olarak, paranın değeri artar ve mal ve hizmetlerin fiyatı düşer. Bu, bireyler ve işletmeler için satın alma gücünün ve yaşam standartlarının artmasına neden olabilir.

Sonuç olarak, CeFi ve DeFi sistemleri, her sistemin enflasyon oranını etkileyebilecek farklı özellik ve nitelikler sergiler. CeFi sistemleri, kısmi rezerv bankaları aracılığıyla enflasyona yol açabilecek yeni para birimleri yaratma yeteneğine sahiptir. DeFi sistemleri, deflasyona yol açabilecek stablecoin'ler oluşturma yeteneğine sahiptir. Bu farklılıkları ve bunların finansal sistemi ve bir bütün olarak ekonomiyi nasıl etkilediğini anlamak önemlidir.

### 3.11. Mixer ve DeFi'de Kara Para Aklama

Mixer ve DeFi kara para aklama, çeşitli aracılar veya platformlar aracılığıyla para aktararak yasa dışı fonların kaynağını gizlemek için kullanılan bir yöntemdir. Bu, yasa dışı fonları takip etmek ve yetkililerin fonların orijinal kaynağını takip etmesini zorlaştırmak amacıyla yapılır.

Kafa karıştırıcı kara para aklama, fonların kaynağını gizlemek için tumbler olarak da bilinen kripto para karıştırıcıları kullanma uygulamasıdır. Bu karıştırıcılar, birçok küçük işlemi alıp karıştırarak çalışır ve bu da orijinal finansman kaynağının belirlenmesini zorlaştırır. Bu teknik genellikle kullanılanlar gibi diğer yöntemlerle birleştirilir. Yetkililerin kafasını daha fazla karıştırmak için birden fazla cüzdan kullanmak veya hileli işlemler oluşturmak. Öte yandan DeFi Kara Para Aklama, yasa dışı fonların kaynağını aktarmak ve gizlemek için DeFi platformlarını kullanır. Bu platformlar, merkezi olmayan ağlar üzerinde çalışır ve kullanıcılarına genellikle anonimlik sunarak bireylerin tespit edilmeden kolayca para göndermesine ve kara para aklamasına olanak tanır (Hinteregger ve Haslhofer, 2018).

Kripto paralar, temel olarak akıllı sözleşmeler ile belirlenmiş özelliklere sahiptir. Belirlenmiş özellikler aracılığı ile kara para aklama süreçlerinin dışında eğer destekleyici kuruluşları tarafından ilk kodlama zamanında ya da çatallanma ile önlem alabilme imkânına sahiptir. Kimi stablecoin'ler hali hazırda kara para aklama ve benzeri durumlar için gelecekte oluşabilecek yasal düzenlemelere hazırlıklı olmak amacıyla akıllı sözleşmelerine kara listeye alma özelliği eklemiştir. Kara listeye alınmış bir cüzdan bir başka cüzdana aktarım yapamamakta ve bu cüzdana gönderimler engellenmektedir. Tablo 2'de görüleceği üzere piyasayı domine eden stablecoin'ler ihtiyaç halinde istedikleri cüzdanın faaliyetini durdurabilmekte ve yapılmış bir varlık transferini iptal edebilmektedir (Paxos, t.y). Stablecoinlerin bir kısmı, resmi web sitelerinde kara listeye alma ve varlık dondurma süreçlerini aktif olarak işlettikleri belirtilmektedir. USDC tarafından hali hazırda 420 adet cüzdan kara listeye alınmıştır (Centre, 2022).

**Tablo 2. Stablecoin'lerin Kara Parayı Aklama Engellemeye Uyumu**

| İsim | İhraçcı | Blok zinciri | KYC gerekliliği | İhraçcı dururabilir mi? | İhraçcı geri kurtarma |
|---|---|---|---|---|---|
| Tether | iFinex, DigFinex | Bitcoin, Ethereum, EOS, Liquid, Tron | Hayır | Evet | Evet |
| USD coin | Coinbase, Circle | Ethereum | Hayır | Evet | Evet |
| PAX | Paxos Trust Company | Ethereum, Binance zinciri | Hayır | Evet | Evet |
| Dai | MakerDAO | Ethereum | Hayır | Hayır | Hayır |
| Binance USD | Paxos Trust Company | Ethereum, Binance zinciri | Hayır | Evet | Evet |

**Kaynak:** URL 5

Kripto para birimleri ve merkezi olmayan platformların kullanımı daha yaygın hale geldikçe, hem mixerler hem de DeFi kara para aklama, dünya çapındaki kolluk kuvvetleri için artan bir endişe kaynağı haline gelmektedir. Yetkililer, bu teknolojilerle mücadele etmek için aşağıdakiler de dahil olmak üzere çeşitli önlemler aldı. Bu faaliyetlere katılanları belirlemek ve izlemek için gelişmiş analitik ve gözetim teknolojisi kullanılabilmektedir (Miers, Garman, Green ve Rubin, 2013).

### 3.12. CeFi ve DeFi Sistemlerinde Güvenlik ve Gizlilik

CeFi, merkezi bir otorite tarafından düzenlenen ve işletilen geleneksel finansal kurumları ifade eder. Bu kurumlar, geleneksel finansal düzenlemelere ve prosedürlere tabi olan bankaları, kredi birliklerini ve diğer finansal kurumları içerir.

DeFi ise blockchain teknolojisine dayalı merkezi bir otorite olmadan işleyen finansal sistemleri ifade etmektedir. DeFi sistemleri, finansal işlemleri ve operasyonları kolaylaştırmak için akıllı sözleşmelere ve merkezi olmayan ağlara güvenmektedir.

Güvenlik ve gizlilikle ilgili olarak, CeFi kurumları genellikle müşteri verilerini ve finansal varlıkları korumak için sağlam güvenlik önlemleri uygulamaktadır. Bu önlemler, gelişmiş şifreleme teknolojisi, güvenli sunucular ve kilitli kasalar ve güvenlik personeli gibi fiziksel güvenlik önlemlerini içermektedir. Ancak, bu kurumlar siber saldırılara ve diğer dolandırıcılık biçimlerine karşı hâlâ savunmasızdır (Miers vd., 2013).

DeFi sistemi ise merkezi olmayan çalışması nedeniyle daha yüksek düzeyde güvenlik ve gizlilik sunmaktadır. Blok zincirindeki işlemler, bilgisayar korsanlarının verileri değiştirmesini veya kurcalamasını zorlaştıran bir genel deftere kaydedilmektedir. Ek olarak, DeFi şemaları genellikle müşteri verilerini ve varlıklarını korumak için gelişmiş şifreleme teknikleri kullanılmaktadır. Ancak DeFi sisteminin kendi güvenlik riskleri de yok değil. Örneğin akıllı sözleşmeler, bilgisayar korsanlarının yararlanabileceği güvenlik açıkları içerebilir. Ek olarak, DeFi sistemleri, CeFi kurumlarıyla aynı düzeyde düzenleyici gözetime sahip olmayabilir, bu da onları dolandırıcılığa ve diğer suistimal biçimlerine karşı daha savunmasız hale getirir (Hinteregger ve Haslhofer, 2018).

Özetle, hem CeFi hem de DeFi sistemlerinin kendi güvenlik ve gizlilik sorunları vardır. CeFi kurumları yüksek düzeyde güvenlik sunar, ancak yine de siber saldırılara ve diğer dolandırıcılık biçimlerine karşı savunmasızdır. DeFi sistemleri, merkezi olmayan operasyonlar nedeniyle daha yüksek düzeyde güvenlik ve gizlilik sunar, ancak aynı zamanda akıllı sözleşme güvenlik açıkları ve düzenleyici gözetim eksikliği risklerine de maruz kalır.

### 3.12.1. CeFi Güvenliği

CeFi sistemleri, finansal verileri depolamak ve işlemek için merkezi sunuculara ve veri tabanlarına güvenir. Bu, merkezi sunucunun ele geçirilmesi veya hacklenmesinin tüm sistemin güvenliğini tehlikeye atabileceği tek bir hata noktasına karşı savunmasız olduğu anlamına gelir. Ek olarak, CeFi sistemleri genellikle işlemleri kolaylaştırmak için güvenlik riskleri oluşturabilecek üçüncü taraf aracılara güvenir.

Örneğin, JPMorgan Chase veri ihlali sonrası 83 milyon müşteri hesabını saldırganların eline geçti. Bu, CeFi sistemlerinin siber saldırılara ve veri ihlallerine karşı savunmasızlığını vurgular (Roman, 2014).

Bu risklere rağmen, CeFi sistemleri genellikle bu tehditlere karşı koruma sağlamak için şifreleme, güvenlik duvarları ve çok faktörlü kimlik doğrulama gibi sağlam güvenlik önlemlerine sahiptir. Ayrıca, Finans Sektörü Düzenleme Kurumu (FINRA) ve Menkul Kıymetler ve Borsa Komisyonu (SEC) gibi düzenleyiciler, güvenlik ve gizlilik standartlarına uyumu sağlamak için CeFi sistemlerini denetler ve düzenler (Manning, 2022).

### 3.12.2. DeFi Güvenliği

DeFi sistemleri merkezi olmayan ağlarda çalışır, yani sistemi kontrol eden merkezi bir otorite veya aracı yoktur. Bunun yerine, işlemler, bunları bir blok zinciri gibi dağıtılmış bir defterde doğrulayan ve kaydeden bir düğüm ağı tarafından işlenir.

Bu merkezi olmayan yapı, çeşitli güvenlik avantajları sunar. Örneğin, DeFi sistemleri tek bir arıza noktasına dayanmadığından, merkezi bir sunucuyu veya veritabanını hedef alan saldırılara karşı daha az savunmasızdır. Ek olarak, DeFi sistemleri, işlemlerin güvenliğini sağlamak için kriptografik algoritmalar kullanır ve bu da onları kurcalamaya veya dolandırıcılığa karşı dirençli kılar.

Ancak DeFi sistemleri güvenlik risklerine karşı bağışık değildir. Örneğin, genellikle DeFi platformlarında işlemleri kolaylaştırmak için kullanılan akıllı sözleşmeler, saldırganlar tarafından istismar edilebilecek güvenlik açıkları içerebilir. Ek olarak, DeFi platformları, kullanıcı kimlik bilgilerini veya fonlarını çalmayı amaçlayan kimlik avı saldırılarına veya diğer sosyal mühendislik biçimlerine karşı savunmasız olabilir (PeckShield, 2020).

### 3.12.3. DeFi ve Gizlilik

DeFi sisteminin ana avantajlarından biri, kullanıcılarına sağladığı anonimliktir. İşlemler dağıtılmış bir deftere kaydedilir, bu da işlemde yer alan tarafların belirlenmesini zorlaştırır. Bu, mali faaliyetlerini gizli tutmayı tercih eden kullanıcılar için gizlilik avantajları sağlayabilir. Ama bu anonimlik iki tarafı keskin bir kılıç da olabilir. Gizlilik avantajları sağlayabilse de kolluk kuvvetlerinin yasa dışı faaliyetleri izlemesini veya DeFi platformlarında düzenlemeleri uygulamasını zorlaştırabilir (Elliptic, t.y.; Chainanalysis, 2019).

### 3.13. CeFi ve DeFi Pazar Manipülasyonu

DeFi ve CeFi piyasaları, son yıllarda birçok yatırımcının finansal hizmetler ve getiri fırsatları için bu platformlara yönelmesiyle önemli bir büyüme yaşadı. Bununla birlikte, diğer tüm pazarlar gibi, hem CeFi hem de DeFi manipülasyona karşı savunmasızdır.

Merkezi finansal platformlar, finansal işlemleri ve karar vermeyi kolaylaştırmak için geleneksel finansal kurumlara veya merkezi kurumlara güvenir. Bu nedenle, içeriden bilgi sızmasına dayalı ticaret ve hileli ticaret gibi geleneksel piyasa manipülasyonlarına karşı savunmasızdır. Yatırım kararları vermek için halka açık olmayan bilgileri kullanan içeriden bilgi ticareti, finansal kurumların çalışanları veya içeriden kişiler piyasayı etkileyebilecek gizli bilgilere erişim sağladığında gerçekleşebilir. Öte yandan, yıkama ticareti, artan piyasa faaliyeti görüntüsü yaratmak amacıyla menkul kıymetlerin alım satımını ifade eder. Her iki manipülasyon türü de yapay olarak fiyatları yükseltebilir veya hatalı talep yaratarak yatırımcıların zarar görmesine neden olabilir (Elliptic, t.y.; Chainanalysis, 2019).

Geleneksel işlem yöntemlerine ek olarak, CeFi piyasası ayrıca algoritmalar ve yüksek frekanslı ticaret (HFT) kullanan manipülasyonlara karşı savunmasızdır. Algoritmik ticaret, önceden belirlenmiş kurallara dayalı bir bilgisayar programı kullanarak ticaret yapmak anlamına gelirken, HFT, işlemlerin genellikle milisaniyeler içinde yüksek hızda yürütülmesini içerir. Bu yöntemler piyasa etkinliğini ve likiditesini artırabilirken, sahte emirler ve "dolandırıcılık" yoluyla piyasa istikrarsızlığı ve manipülasyonu da yaratabilir (Zhou, Qin, Torres, Le ve Gervais, 2021; Qin, Zhou ve Gervais, 2022).

CeFi piyasalarının aksine, DeFi platformları merkezi değildir ve işlemleri ve karar almayı kolaylaştırmak için blockchain teknolojisini kullanır. Bu ademi merkeziyet şeffaflığı ve güvenliği artırırken, piyasa manipülasyonu riskini tamamen ortadan kaldırmaz. DeFi piyasasında meydana gelen bir tür manipülasyon, "halı çekme"dir. Bu, projenin geliştiricisinin veya ekibinin, yatırımcının parasıyla projeyi aniden sonlandırması anlamına gelir. Bu, yatırımcılar için önemli kayıplara neden olabilir ve ayrıca bir bütün olarak DeFi piyasasına olan güvenin kaybolmasına neden olabilir (Qin vd., 2022).

DeFi piyasasında meydana gelen başka bir manipülasyon biçimi de "önden çalıştırma"dır. Bu, yaklaşan ticaret hakkındaki bilginizi kullanarak orijinal tüccarın önüne bir ticaret yapma eylemini ifade eder. Bu, alıcılar ve satıcılar arasındaki hüküm ve koşulların doğrudan kod satırlarına yazıldığı, kendi kendini yürüten anlaşmalar olan akıllı sözleşmeler kullanılarak yapılabilir. Akıllı sözleşmeler verimliliği ve güvenliği artırabilir ancak kötü niyetli aktörler tarafından da kötüye kullanılabilir (Daian vd., 2019).

Hem CeFi hem de DeFi piyasaları, piyasa manipülasyonu sorununu çözmek için çeşitli önlemler almıştır. CeFi piyasalarında, Menkul Kıymetler ve Borsa Komisyonu (SEC) ve Finans Sektörü Düzenleme Kurumu (FINRA) gibi düzenleyiciler, içeriden öğrenenlerin ticaretini ve diğer manipülasyon biçimlerini önlemek için kurallar ve düzenlemeler uygular. Projeleri ve platformları denetlemek ve en iyi uygulamaları oluşturmak için DeFi pazarında merkezi olmayan borsalar (DEX'ler) ve özdenetim kuruluşları (SRO'lar) kurulmuştur (Manning, 2022).

Ek olarak, yapay zeka (AI) ve makine öğrenimi algoritmaları, hem CeFi hem de DeFi pazarlarında potansiyel olarak piyasa manipülasyonunu tespit edip önleyebilir. Örneğin, yapay zekâ, piyasa verilerini analiz etmek ve yetkilileri potansiyel manipülasyona karşı uyarmak için olağandışı kalıpları ve davranışları belirlemek için kullanılabilir.

### 3.14. CeFi ve DeFi Ortak Çalışabilirliği

Araştırmacılar ve uygulayıcılar tarafından belirlenen CeFi ve DeFi arasında birkaç ortak çalışabileceği ve birbirine olumlu yönde katkı sunabileceği nokta bulunmaktadır. Birinci, DeFi'nin, temel ağların merkezi olmayan yapısı nedeniyle CeFi'ye kıyasla potansiyel olarak daha yüksek düzeyde şeffaflık ve güvenlik sunabilmesidir. Bu, DeFi'yi, tek bir başarısızlık noktası riski veya dolandırıcılık veya kötü yönetim potansiyeli gibi merkezileştirme risklerinden endişe duyan kişi ve kurumlar için daha çekici hale getirebilir.

Diğer bir nokta ise DeFi'nin coğrafya, gelir veya diğer faktörler nedeniyle geleneksel finansal sistemlerden dışlanabilecek bireylere ve topluluklara finansal hizmetlere erişim sağlayarak potansiyel olarak daha fazla finansal katılım sunabilmesidir. DeFi ayrıca potansiyel olarak alternatif finansal araçlara ve kripto para birimi ve diğer dijital varlıklar gibi geleneksel finansal kanallar aracılığıyla kullanılamayan varlık sınıflarına daha fazla erişim sağlayabilir.

Son olarak, DeFi, belirli süreçlerin otomasyonu ve aracılara ihtiyaç duymadan çalışabilmesi nedeniyle CeFi'ye kıyasla potansiyel olarak daha fazla verimlilik ve maliyet tasarrufu sunabilir. Bu, DeFi'yi hem finansal hizmet kullanıcıları hem de sağlayıcıları için daha çekici hale getirebilir.

### 3.14.1. Blok Zincir Köprüleri

Blockchain köprüleri, farklı blockchain ağları arasında birlikte çalışabilirliği sağlayan teknolojiyi ifade eder. Başka bir deyişle, farklı blockchain sistemleri arasında bilgi, varlık ve diğer bilgilerin transferine izin verirler.

Blockchain köprülerini düşünmenin bir yolu, onları farklı blockchain dilleri arasında "çeviri hizmetleri" olarak düşünmektir. Nasıl ki dünyada farklı diller konuşuluyorsa, farklı blockchain ağları oluşturmak için farklı protokoller ve teknolojiler kullanılıyor. Bu diller arasında çeviri yapma yeteneği olmadan birbirleriyle iletişim kurmaları ve bilgi alışverişinde bulunmaları zordur (Ellis, Juels ve Nazarov, 2017; Synthetix, t.y.).

Blockchain köprüsüne bir örnek, farklı blockchain ağları arasında veri aktarımını kolaylaştırmak için Interledger Protocol (ILP) adlı bir teknolojiyi kullanan Cosmos Network'tür. Bu, blockchain teknolojisini kullanırken daha fazla esneklik ve ölçeklenebilirlik sağlayan merkezi olmayan bir ağ oluşturulmasına izin verir (Cosmos, t.y).

Başka bir örnek, Ethereum ve Ethereum Classic blockchain ağları arasında varlıkların transferini kolaylaştırmak için Parity Bridge adlı bir teknolojiyi kullanan Ethereum tabanlı POA ağıdır (POA, t.y).

Genel olarak, blok zinciri köprülerinin kullanımı, blok zinciri teknolojisinde önemli bir gelişmedir çünkü farklı blok zinciri sistemleri arasında daha iyi birlikte çalışabilirlik ve bağlantı sağlar. Bu, blockchain teknolojisinin kullanımında daha fazla verimlilik ve ölçeklenebilirliğe yol açabilir ve nihayetinde benimsenmesine ve genel kullanımına katkıda bulunabilir.

### 3.14.2. DeFi: CeFi'ye Yenilikçi Bir Bakış Açısı

Merkezi Olmayan Finans'ın kısaltması olan DeFi, blockchain teknolojisi gibi merkezi olmayan ağlarda çalışan finansal sistemleri tanımlamak için kullanılan bir terimdir. DeFi, bankalar gibi merkezi kurumlar aracılığıyla faaliyet gösteren geleneksel finansal sistemleri ifade eden CeFi'ye yenilikçi bir katkı olarak kabul edilir.

DeFi ve CeFi arasındaki temel farklardan biri, DeFi'nin merkezi olmayan ağlarda çalışması, yani tek bir varlık veya grup tarafından kontrol edilmemesidir. Bu merkezsizleştirme, CeFi sistemlerine kıyasla daha fazla şeffaflık, erişilebilirlik ve güvenlik sunar.

DeFi, borç alma, borç verme, ticaret ve ödemeler gibi çeşitli finansal hizmetleri içerir. Alıcı ve satıcı arasındaki sözleşmenin şartlarının doğrudan bir kod satırına yazıldığı, kendi kendini yürüten sözleşmeler olan akıllı sözleşmeler de kullanılır.

DeFi'nin avantajlarından biri, bankalar ve kredi kurumları gibi geleneksel aracılara ihtiyaç duymadığı için daha fazla finansal kapsama izin vermesidir. Bu, konumu veya kredi geçmişi nedeniyle geleneksel finansal hizmetlere erişimi olmayan kişilerin yine de DeFi'ye katılabileceği anlamına gelir.

DeFi, geleneksel finansal sistemlerde bulunmayabilecek yeni finansal araçların ve piyasaların yaratılmasına olanak sağladığı için yüksek kar potansiyeli nedeniyle de popülerlik kazanmıştır. DeFi protokolleri CeFi hizmetlerini kopyalamak olarak değil yenilikçi bakış açıları geliştirmektedir. Örnek olarak otomatik piyasa yapıcı protokollerinin gelişi alım satım kayıtlarının tutulması ve aktarımları merkezi çözümlerden daha verimli hale getirmiştir. Bununla birlikte, DeFi hala nispeten yeni ve hızla gelişen bir sektör ve bu nedenle yatırım yapmadan önce dikkatlice değerlendirilmesi gereken belirli riskler ve belirsizlikler taşıyor (Bank of International Settlements, 2020; Zhou vd., 2021; Binance, 2020).

## 4. SONUÇ

Bu çalışmada, CeFi ve DeFi arasında sunduğu hizmetler açısından kayda değer bir farklılık olmadığı görülmüştür. DeFi platformlarında kullanıcıların CeFi'nin aksine yasal olarak koruma altından olmadığı anlaşılmaktadır. Her iki finans alanında temel ayrışmanın, varlık kontrolünde yetkinin tüm süreç boyunca kimde olduğu ve süreçlerin şeffaflığı faktörlerinde gerçekleştiği tespit edilmiştir. DeFi platformlarının yasal mevzuata henüz uygun olmadığı ancak CeFi temelinde çalışan ancak DeFi projeleri ile etkileşimde olan projelerin KYC sağlamaktadır. DeFi'nin kurumsal firmalar tarafından benimsenmesinin önündeki en büyük engellerden biri kara para aklama süreçlerine dahil olabilmesidir. Ancak bazı stable coinlerin yasal düzenlemelere uyumlu olabilmek adına ve gelecekte ortaya çıkabilecek AML problemlerine hazırlıklı olmak için akıllı sözleşmelerinde cüzdanların kara listeye alınması özelliği kodlanmıştır. Tether, BUSD, USDC gibi stablecoinlerin AML ile uyumlu olduğu anlaşılmıştır.

**KAYNAKÇA**